\documentclass[prb,twocolumn]{revtex4-1} 

\usepackage{hyperref}  
\usepackage{amsmath}  
\usepackage{amsfonts} 
\usepackage{graphicx} 

\begin{document}


\title{An adaptive gaussian quadrature for the Voigt function}

\author{Fr\'ed\'eric Paletou} \email{frederic.paletou@univ-tlse3.fr}
\affiliation{Universit\'e de Toulouse, Observatoire
  Midi--Pyr\'en\'ees, Cnrs, Cnes, Irap, F--31400 Toulouse, France}

\author{Christophe Peymirat} \email{christophe.peymirat@univ-tlse3.fr}
\affiliation{Universit\'e de Toulouse, Facult\'e des Sciences et
  d'Ing\'enierie, F--31062 Toulouse cedex 9, France}

\author{\'Eric Anterrieu} \email{eric.anterrieu@cesbio.cnes.fr}
\affiliation{Cesbio, Cnrs UMR 5126, Universit\'e de Toulouse, Cnes,
  Ird, F--31401 Toulouse, France}

\author{Torsten B\"{o}hm} \email{tboehm@irap.omp.eu}
\affiliation{Irap, Cnrs UMR 5272, Universit\'e de Toulouse, Cnes,
  F--31400 Toulouse, France}


\date{\today}

\begin{abstract}
  We evaluate an adaptive gaussian quadrature integration scheme that
  will be suitable for the numerical evaluation of generalized
  redistribution in frequency functions. The latter are indispensable
  ingredients for ``full non--LTE'' radiation transfer computations
  i.e., assuming potential deviations of the velocity distribution of
  massive particles from the usual Maxwell--Boltzmann distribution. A
  first validation is made with computations of the usual Voigt profile.
\end{abstract}

\maketitle 

\section{Introduction}

Radiation transfer is, by essence, a difficult problem (e.g., Rutily
\& Chevallier 2006), as well as a question of very large relevance in
astrophysics. It relies indeed on \emph{complex} non-linear
light--matter interactions (see e.g., Hubeny \& Mihalas 2014, Rutten
2003).

At the very heart of the problem lays the issue of how photons may
scatter on these {\em moving} massive particles constituting the
atmosphere under study. The usual literature classify these processes
as, either ``complete'' redistribution in frequency (CRD), or
``partial'' redistribution in frequency (PRD; see e.g., \S10 of Hubeny
\& Mihalas 2014). The vast majority of astrophysical problems are
solved, still, within the frame of CRD, for which further
simplifications are the equality of emission and absorption profiles
-- the latter being also usually known {\em a priori} --, which also
leads to the independence of the so-called source function
vs. frequency.

Besides, and more generally, while non-equilibrium distributions of
photons i.e., potential departures from the Planck law, have been
routinely considered since the late 60's, a very limited number of
studies tried to push further the description of the physical problem,
by questioning {\em to what extent the most often assumed
  Maxwell--Boltzmann velocity distribution of the massive particles
  onto which photons scatter may remain valid} (see e.g., Oxenius \&
Simmonneau 1994, and references therein).

Non-maxwellian velocity distributions functions (hereafter vdf) have
been studied (e.g., Scudder 1992 and further citations) or evidenced
in \emph{natural} plasma (see e.g., Jeffrey et al. 2017 for a recent
study about solar flares). Such departures from Maxwell--Boltzmann
vdf's have also been considered in the radiative modelling of spectral
lines formed in neutral planetary exospheres (e.g., Chaufray \&
Leblanc 2013), where these authors introduced so-called $\kappa$ vdf's
into their photon scattering physical model.

However, such non-maxwellian vdf's are still known {\em ab initio}
before solving the radiation transfer problem. The more general issue
of computing {\em self-consistent} non-equilibrium distributions for
both photons and massive particles -- whose associated problem we
coin ``full non--LTE radiation transfer'' -- remains a quite open
question in astrophysics, although a few studies have already been
conducted in the past (see e.g., Borsenberger et al. 1986, 1987).

Hereafter, we provide a first numerical tool that will allow us
to go further in this direction, enabling further computations of {\em
  generalized} redistribution functions. Moreover the numerical scheme
we evaluated may also be of more general interest, for other topics of
numerical (astro)physics.

\section{Redistribution in frequency}

As an illustrative but important example, we shall focus here on the
case of coherent scattering in the {\em atomic} frame of reference,
for a spectral line of central wavelength $\nu_0$. We shall also
assume that only ``natural'' broadening is at play for the upper
energy level of, typically, a resonance line with an infinitely sharp
lower level.

Therefore, we shall consider an elementary {\em frequency}
redistribution function $r(\xi',\xi)$ such that:

  \begin{equation}
    r(\xi',\xi) = \varphi(\xi') \delta(\xi'-\xi) \, ,
  \end{equation}
where $\xi'$ and $\xi$ are, respectively, the incoming and the
outgoing frequencies of a photon, and $\delta$ is the usual Dirac
distribution, together with:

  \begin{equation}
    \varphi(\xi') = \left( \frac{\Gamma}{\pi} \right) {\frac{1}{(\xi' -
        \nu_0)^2 + {\Gamma}^{2}}}  \, .
  \end{equation}
The latter is a {\em Lorentzian} profile, with damping parameter $\Gamma$,
resulting from the ``natural width'' of the upper atomic state of the
transition at $\nu_0$.

If we assume that the {\em angular} redistribution associated with the
scattering event is {\em isotropic}, such a case of radiation damping
and coherence in the atom's frame refers to the standard case ``II-A''
in the nomenclature of Hummer (1962; see also Hubeny \& Mihalas 2014).

Once the elementary scaterring process have been defined in the atomic
frame of reference, we have to consider for a further practical
implementation into a radiative transfer problem, the collective
effects induced by the agitation of a pool of massive particles
populating the atmosphere. This is precisely in this ``jump'' to the
observer's frame of reference, because of Doppler shifts such as:

\begin{equation}
    \nu = \xi + \frac{\nu_0}{c} {\vec{n} . \vec{v}}  \, ,
\end{equation}
where $\nu$ is the observed frequency, $\vec{n}$ may be either the
incoming or the outgoing direction of a photon, and $\vec{v}$ the
velocity of the massive particle onto which the scattering takes
place, that some assumption has to be made about the vdf of the
massive atoms (or molecules) present in an atmosphere, under given
physical conditions.

Detailed derivations of $R_{II-A}$ can be found in the classical
literature about redistribution functions, from Henyey
(1940)\cite{typos} to Hummer (1962).  Standard redistribution
functions have been first derived assuming that the vdf of the atoms
scattering light is a Maxwell--Boltzmann distribution. Then, but more
generally, any macroscopic redistribution function in the observer's
frame suitable for implementation into the numerical radiative
transfer problem will result from the further integration along each
velocity components $u_i$ (hereafter normalized to the most probable
velocity $v_{\rm th.}=\sqrt{2kT/m}$) characterizing the movement of
the scattering atoms and, therefore considering these changes of
frequencies due to the associated Doppler shifts as expressed by
Eq.\,(3). The latter phenomenon is usually refered to as Doppler, or
{\em thermal} broadening.
  
\section{The numerical problem}

\begin{figure}[]
  \includegraphics[width=9. cm,angle=0]{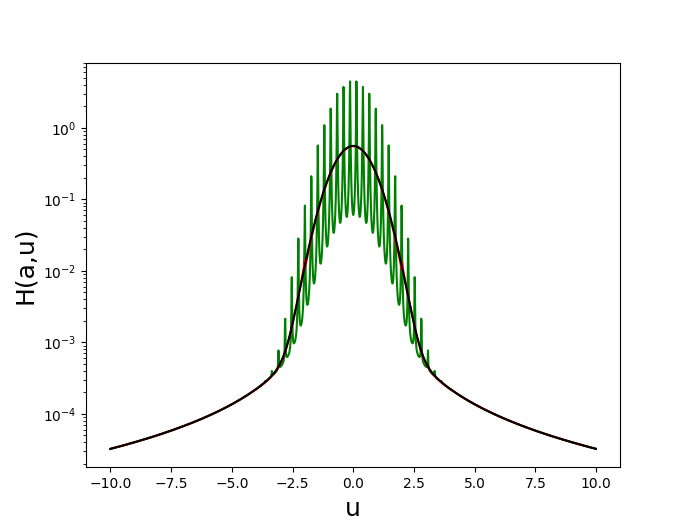}
  \caption{The failure of a standard Gauss--Hermite quadrature of order
    $k=70$ (green), as compared to the almost superimposed results
    from, respectively, the method using the Faddeeva complex function
    (dark) and our alternative double adaptive Gaussian quadrature
    scheme, for a normalized Voigt profile with $a=10^{-2}$.}
  \label{Fig1}
\end{figure}

We aim at generalizing computations of redistribution functions in
order to be able to compute vdf's self-consistently with the radiation
field. Therefore we need a \emph{robust} numerical approach to repeatedly
perform numerical integrations like:

\begin{widetext}
\begin{equation}
    H_{1}(x',x,\gamma) = \displaystyle \int_{-\infty}^{+\infty} { {
        {f(u_1)du_1} \over { [ (\frac{x+x'}{2}) \sec(\gamma/2) - u_1
          ]^2 + [ \frac{a}{{\Delta \nu_{D}}} \sec(\gamma/2)]^2 } } }
    \, ,
  \end{equation}
\end{widetext}
where $x'$ and $x$ are the usual incoming and outgoing {\em
  reduced}\cite{redfreq} frequencies in the observer's frame, ${\Delta
  \nu_{D}}$ the Doppler width defined as $(\nu_0/c)v_{\rm th.}$, and
$\gamma$ the diffusion angle between incoming and outgoing directions
in the plane defined by $u_1$ and $u_2$. For the Maxwell--Boltzmann
case, we should indeed use:

\begin{equation}
  f(u_1)=\frac{1}{\sqrt{\pi}} e^{-u_1^2}  \, ,
\end{equation}
but we shall need to consider $f(u_1)$ to be $non-$analytic, and, at
first, (slightly) departing from the maxwellian standard vdf. Indeed,
physical conditions leading to small departures from a Gaussian vdf
have already been identified and discussed by Oxenius (1986), and they
would correspond to a non--LTE gas of moderate optical thickness. Note
also that, for a preliminary study, we shall assume a self-consistent
vdf solution of the problem that may still be decomposed as
$f_1(u_1)f_2(u_2)f_3(u_3)$.

However, before exploring potential departures from gaussianity, we
need to adopt a robust enough numerical strategy in order to
numerically evaluate integrals such as Eq.\,(4), a task which is
notoriously difficult even with Maxwell--Boltzmann fdv's. It is very
easy to verify that, for instance a standard Gauss--Hermite (GH)
quadrature, even at high rank $k$, fails at computing properly a
somewhat simpler expression like the Voigt\cite{voigt} function given
in Eq.\,(13). We display in Fig. (1) the comparison between a GH
integration and the new numerical scheme which is presented hereafter.

\section{Adaptive Gaussian Quadrature}

We shall start following the scheme proposed by Liu \& Pierce (1994),
which is based on the classical Gauss--Hermite (GH) quadrature. The
latter is indeed suitable for integrations of the kind:

\begin{equation}
    I=\displaystyle \int_{-\infty}^{+\infty} { f(y){ {e^{-y^2}dy} } }  \, .
  \end{equation}
Then the GH quadrature is such that:

\begin{equation}
  \displaystyle \int_{-\infty}^{+\infty} { f(y){ {e^{-y^2}dy} } }
  \simeq \sum_{i=1}^{k} {w_i f(y_i)} 
\end{equation}
where the nodes $y_i$ are the zeros of the $k$-th order Hermite
polynomial, and $w_i$ the corresponding weights. Tabulated values of
both nodes and weights can be found very easily, and they are also
available for various programming language. We shall use {\tt numpy}'s
(Oliphant 2006) function {\tt polynomial.hermite.hermgauss}, and a GH
of order $k=70$ for all results presented hereafter.

The main drawback of such a standard quadrature is that function $f$
shall be scanned at the very nodes $y_i$ {\em irrespectively from} the
range where it may have its most significant variations.

However, Liu \& Pierce (1994) proposed that, should a function $g$ to
be integrated, one may define:

\begin{equation}
  h(y)= { {g(y)} \over {{\cal N}(y;\hat{\mu},\hat{\sigma})} }  \, ,
\end{equation}
where ${\cal N}$ is the usual Gaussian function:

\begin{equation}
  {\cal N}(y;\hat{\mu},\hat{\sigma}) = \frac{1}{\hat{\sigma} \sqrt{2\pi}}
    e^{-\frac{1}{2} (\frac{y-\hat{\mu}}{\hat{\sigma}})^2 }    \, ,
\end{equation}
so that one can write:

\begin{equation}
  \displaystyle \int_{-\infty}^{+\infty} { g(y) dy } = \displaystyle
  \int_{-\infty}^{+\infty} { h(y){{\cal N}(y;\hat{\mu},\hat{\sigma})}
    dy } \, ,
\end{equation}
and, finally:

\begin{equation}
  \displaystyle \int_{-\infty}^{+\infty} { g(y) dy } \simeq
  \sum_{i=1}^{k} {{{w_i} \over {\sqrt{\pi}}} h(\hat{\mu}+ \sqrt{2}
    \hat{\sigma} y_i)}  \, .
\end{equation}
This adaptive Gaussian quadrature scheme (AGQ) allows to use the
original nodes and weights of the GH quadrature, but somewhat {\em
  zooms in} these domains where function $g$ has its most
significant variations.

The choice of $\hat{\mu}$ and $\hat{\sigma}$ is of importance. Liu \&
Pierce (1994) suggested to adopt $\hat{\mu}$ to be the mode of $g$,
and $\hat{\sigma}=1/\sqrt{j}$, where:

\begin{figure}[]
  \includegraphics[width=9. cm,angle=0]{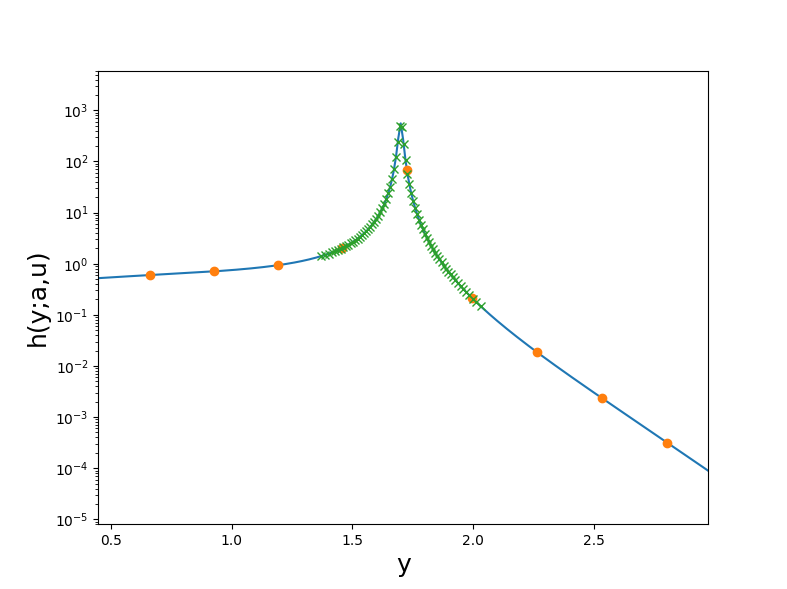}
  \caption{Example of distribution of nodes for an initial
    Gauss-Hermite quadrature of order $k=70$ (dots), and for our
    adaptive Gaussian quadrature centered at the Lorentzian peak
    (crosses).}
  \label{Fig2}
\end{figure}

\begin{equation}
    j = - \frac{\partial^2}{\partial y^2} \log g(\hat{\mu}) \, .
\end{equation}
We shall come back on this choice in the following section, and show
that a somewhat larger $\hat{\sigma}$ value is more suitable for the
special case of the Voigt profile.

\section{AGQ tests with the Voigt function}
  
Let us consider the {\em normalized} Voigt function
hereafter defined as:

\begin{equation}
  H(a,u) = \frac{1}{\sqrt{\pi}} \left( \frac{a}{\pi} \right)
    \int_{-\infty}^{+\infty} { e^{-y^2} dy \over {(u-y)^2 + a^2} } \,
    ,
\end{equation}
and which satisfies to:

\begin{equation}
\int_{-\infty}^{+\infty} {H(a,u) du} = 1 \, .
\end{equation}
Note that several authors use $H$ for the Voigt profile normalized to
$\sqrt{\pi}$, but $U$ instead of our $H$ normalized to unity though
(see e.g., Hubeny \& Mihalas 2014, their \S 8). We shall also use $h(y;
a,u)$ for the integrand of Eq.\,(13).

For this numerical integration, three main regimes should be
considered, depending on the values of $u$ i.e., according to the
respective amplitudes of the Gaussian and the Lorentzian components of
the integrand. For the line core region such that $\lvert u \rvert <
2$, we use a slightly modified AGQ, for which we use a value of
$\hat{\sigma}$ larger than the one suggested in the original article
of Liu \& Pierce (1994). We display in Fig.\,(2) the new quadrature
nodes, marked with crosses, centered at the Lorentzian peak located at
$y \approx 1.7$, and using $3\hat{\sigma}$ instead of the value
suggested in the original prescription of Liu \& Pierce (1994). The
nodes of the standard GH quadrature (at the \emph{same} order) are
displayed as dots. They extend too far away, clearly ``miss'' the
large amplitude Lorentzian peak and therefore the dominant
contribution to the integral.

Second, we perform a {\em double} AGQ scheme for the near wing regions
such that $2 < \lvert u \rvert < 4$, and for which two discernable
peaks of \emph{comparable} amplitudes result from, respectively, the
Lorentzian and the Gaussian components of the convolution (we shall
hereafter refer to $u_2$ and $u_4$ for these two boundary values). In
such a case, we use both the centering and integration range controls
provided by the original AGQ for evaluating \emph{separately} the
contribution from each component of the integrand. For the Lorentzian
component we therefore do as when $\lvert u \rvert < 2$, but {\em we
  add} to this part of the integral the contribution of the nearby
Gaussian peak using {\em another} AGQ centered at 0, and of specific
$\hat{\sigma}_G$ obviously adapted to the width of the known Gaussian
component of the integrand (also, overlap with the nearby Lorentzian
component should be avoided). The two distincts sets of nodes, based
on the same original GH quadrature nodes, at same order, are displayed
by crosses of different colors in Fig.\,(3).

\begin{figure}[]
  \includegraphics[width=9. cm,angle=0]{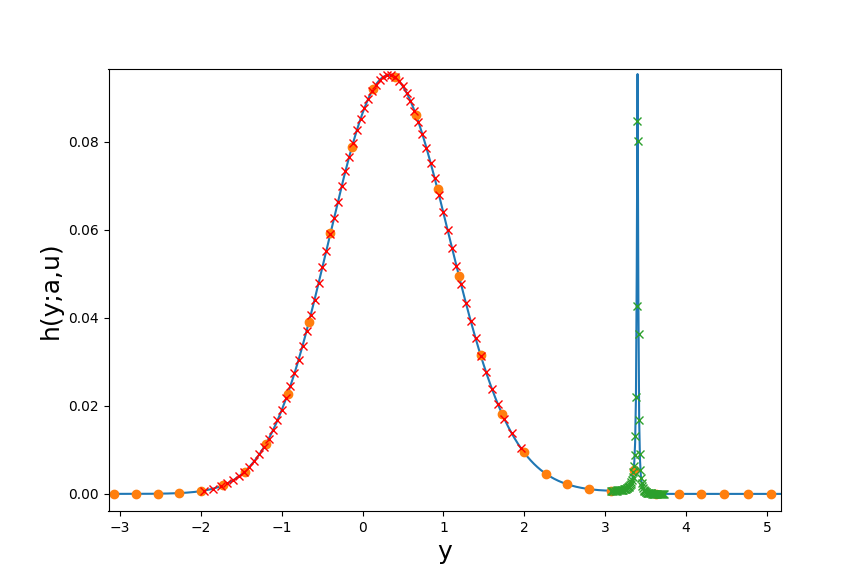}
  \caption{Respective distributions of nodes, marked by crosses of
    different colors, from an initial Gauss-Hermite quadrature of
    order $k=70$ when the Gaussian and Lorentzian peaks are of
    comparable amplitude, here for $u$ around 3.}
  \label{Fig3}
\end{figure}

Finally, for the far wing where $\lvert u \rvert > u_4$, and when the
Lorentzian peaks fade out, the \emph{usual} Gauss--Hermite quadrature is
satisfactory.

\begin{figure}[]
  \includegraphics[width=9. cm,angle=0]{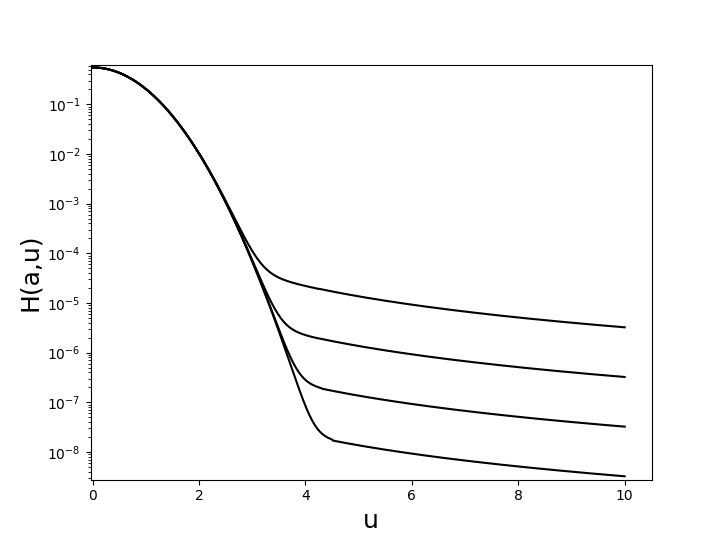}
  \caption{Voigt profiles $H(a,u)$ computed with our double AGQ scheme
    for, respectively, $a=0.001$, $10^{-4}$, $10^{-5}$ and $10^{-6}$
    (and decreasing wing values). Small discontinuities are still
    noticeable at the transitions values about 2 and 4. This should
    however not impair any standard scattering integral computation.}
  \label{Fig4}
\end{figure}

Results using our double AGQ quadrature scheme are displayed in
Fig.\,(4), for different values of $a$ ranging from 0.01 to $10^{-4}$,
more likely regimes expected for our next computations. Maximum
relative errors computations using the Faddeeva function method as a
benchmark, and the {\tt scipy.special.wofz} Python function, are at
most of a few percents, as displayed in Fig.\,(5); note also that the
latter was obtained using a $u_4$ value of 4.25, instead of the
fiducial value of 4, indicating also in what direction a further
fine-tuning could be worked out, if necessary, by considering $u_2$
and $u_4$ as slowly varying functions of $a$.  Sometimes we can still
notice small discontinuities at the changes of regimes, at $u_2$ and
$u_4$. We believe however that, should our procedure be used for Voigt
profile computations and radiative modelling, such small and very
local discontinuities will not impair further computations of these
scattering integrals entering the equations of the statistical
equilibrium.

This new numerical scheme is particularly efficient for {\em small}
values of $a$, typically lower than 0.01, where other schemes may fail
(see for instance the discussion in Schreier 2018 about the
implementation of {\tt Voigt1D} in the {\tt astropy} package in
Python, using the McLean et al. 1994 method). But first of all, it is
certainely suitable for our next applications of such a numerical
integration scheme, and for physical conditions leading to very sharp
Lorentzian peaks. We could also test the sensitivity of our scheme to
the \emph{order} of the initial Gauss--Hermite quadrature. For
instance, for $a<0.01$ we could go down to orders 40 to 50 without any
significant loss of accuracy.

\begin{figure}[]
  \includegraphics[width=9. cm,angle=0]{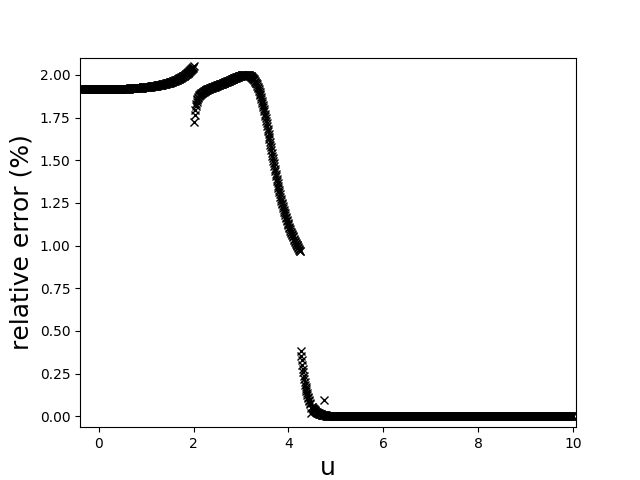}
  \caption{Relative error between our computations with the double AGQ
    method, for $a=10^{-4}$, and a reference computation using the
    Faddeeva complex function.}
  \label{Fig5}
\end{figure}

For larger values of $a$, typically more than 0.1, we noticed that
{\em no} intermediate scheme between the original Liu \& Pierce (1994)
at line core, and the Gauss--Hermite in the wings appears
necessary. However the transition value between the two regimes should
be adapted to the value of $a$, in a 2 to 4 range.

\section{Conclusion}

We have tested a suitable numerical strategy for our first step
towards ``fully non-LTE'' radiative transfer calculations, and the
computation of generalized frequency redistribution functions. We
modified the original strategy of Liu \& Pierce (1994), but also
applied it to a \emph{non}-unimodal distribution.

Our numerical scheme does not pretend to compete with these numerical
methods implemented for the {\em very} accurate computation of the
Voigt function (see e.g., Schreier 2018 and references therein) since
our aim is elsewhere, i.e. to explore departures from Gaussian
vdf's. It is however providing very good results as compared to
reference computations, such as the one using the Faddeeva complex
function. Relative errors down to a few percent are systematically
reported in the near wing region, and we believe that further fine
tuning could be achieved for reaching an even better accuracy.

This is however not the scope of our study, which aims at computing
generalized redistribution functions, after self--consistent
computations of both massive particles and photons respective
distributions under various physical conditions. In that respect, our
main concern is well about a proper ``capture'' of the expected {\em
  very} sharp, and therefore very large amplitude Lorentzian peaks.
And we believe that the principle of our numerical integration scheme
should remain valid for the more easy to track contribution from the
velocity distribution function, even for computed perturbations from a
Gaussian shape.

As a final remark, we are also aware that computations with
\emph{non}-Gaussian functions convolved with a Lorentzian may also be
doable, using a Fourier transform based method (e.g., Mendenhall
2007).

\begin{acknowledgments}
  Fr\'ed\'eric Paletou is grateful to his radiative transfer {\it
    sensei}, Dr. L.H. ``Larry'' Auer, with whom we started discussing
  about these issues long time ago.
\end{acknowledgments}


\begin{thebibliography}{99}
\bibitem{typos} Note that there is a ``typo'' or mistake in this
  article, more specifically in its Eq. (3). A $\cos^2(\alpha/2)$ term
  appears, instead of the correct $\csc^2(\alpha/2)$, where his
  $\alpha$ is our $\gamma$.
\bibitem{redfreq} The usual reduced frequency $x$ is the difference
  between frequency $\nu$ in the observer's frame and the central
  frequency $\nu_0$ of a spectral line, divided by the Doppler width
  ${\Delta \nu_{D}}$.
\end{thebibliography}
\end{document}